\documentclass[11pt,preprint]{aastex}

 \newcommand{\lb}{$\lambda$}

 \slugcomment{Accepted to ApJ.}

 \shorttitle{\ion{O}{1} emission in AGNs}
 \shortauthors{Rodr\'{\i}guez-Ardila et al.}

\begin{document}

 \title{The \ion{O}{1} line emission in Active Galactic Nuclei revisited}

 \author{A. Rodr\'{\i}guez-Ardila\altaffilmark{1,2} and S. M. Viegas}
 \affil{Instituto Astron\^omico e Geof\'{\i}sico - Universidade de S\~ao
 Paulo,
 Av. Miguel Stefano 4200, CEP 04301-904, S\~ao Paulo, SP, Brazil}

 \author{M. G. Pastoriza}
 \affil{Departamento de Astronomia - UFRGS. Av. Bento Gon\c calves 9500,
 CEP 91501-970, Porto Alegre, RS, Brazil}

 \author{L. Prato\altaffilmark{1}}
 \affil{Department of Physics and Astronomy, UCLA, Los Angeles, CA
 90095-1562}

 \and

 \author{Carlos J. Donzelli}
 \affil{IATE, Observatorio Astron\'omico, Universidad Nacional de
       C\'ordoba, Laprida 854, 5000, C\'ordoba, Argentina.}

 \altaffiltext{1}{Visiting Astronomer at the Infrared Telescope facility, which
is operated by the University of Hawaii under contract from the National 
Aeronautics and Space Administration}
 \altaffiltext{2}{e-mail address: ardila@iagusp.usp.br}

  \begin{abstract}
  UV, visible, and near$-$infrared spectroscopy is used to study the
  transitions of neutral oxygen leading to the emission of broad \ion{O}{1}
  $\lambda$8446, $\lambda$11287 and $\lambda$1304 in Active Galactic
  Nuclei. From the strength of the former two lines, contrary to the
  general belief, we found that in six out of seven galaxies,
  L$\beta$ fluorescence is not the only mechanism
  responsible for the formation of these three lines. Because 
  \ion{O}{1} $\lambda$13165 is almost reduced to noise level, 
  continuum fluorescence is ruled
  out as an additional excitation mechanism, but the presence of
  \ion{O}{1} $\lambda$7774 in one of the objects suggests that collisional
  ionization may have an important role in the formation of
  \ion{O}{1} $\lambda$8446. The usefulness of the
  \ion{O}{1} lines as a reliable reddening indicator for the
  broad line region is discussed. The values of E(B-V) derived
  from the \lb1304/\lb8446 ratio agree with those obtained using
  other reddening indicators.  The observations point toward
  a break in the one-to-one photon relation between $\lambda$8446 and
  $\lambda$1304, attributable to several destruction mechanisms that may
  affect the latter line.
  \end{abstract}


  \section{Introduction}

  Permitted transitions of \ion{O}{1} in
  the spectra of gaseous nebulae are very important tracers
  of atomic processes other than recombination, as well as
  indicators of the physical conditions of the emitting gas.
  For many years, \ion{O}{1} $\lambda$8446, usually the strongest
  \ion{O}{1} line in the optical region, has been broadly studied
  in a large number of objects such as the Orion nebula
  \citep{morgan71, grandi75}, planetary
  nebulae \citep{grandi76}, and the pre$-$main-sequence
  Herbig AeBe star LkH$\alpha$ 101 \citep{rudy91}. In active galactic 
  nuclei (AGN), this
  feature is very common. \citet{gran80} presented
  observational data showing that its strength could reach
  several percent of H$\alpha$. He associated the \ion{O}{1} emission
  with a phenomenon restricted to the broad line region (BLR).
  Later, \citet{morward89} showed the lack
  of detectable \ion{O}{1} $\lambda$8446 in Seyfert
  2 galaxies, confirming earlier predictions.

  Usually, resonance fluorescence by Ly$\beta$ has been
  invoked as the single dominant mechanism for the production of
  \ion{O}{1} $\lambda$8446 in AGN since \citet{gran80}.
  This process is favored by a coincidence of energy
  levels between \ion{O}{1} and H\,I: the 2$p~^{3}P-3d~^{3}D^{0}$
  transition of \ion{O}{1} at 1025.77 \AA\ falls within the Doppler
  core of Ly$\beta$ (1025.72 \AA) for gas at 10$^{4}$ K, a
  temperature that is easily reached in the BLR. If this
  process is the dominant one, \ion{O}{1} $\lambda$11287 and
  \ion{O}{1} $\lambda$1304 should also be observed, since they
  are by-products formed during the cascading to the ground
  level following the absorption of a Ly$\beta$ photon.
  In addition, the photon flux ratio between $\lambda$11287
  and $\lambda$8446 should be unity since every $\lambda$11287
  photon leads to the emission of exactly one $\lambda$8446
  photon. Smaller values of this
  ratio should indicate the presence of recombination,
  continuum fluorescence, or other processes that enhance
  $\lambda$8446 but not $\lambda$11287.

  Determining the mechanisms leading to the \ion{O}{1} emission
  in AGN is important for various
  reasons. If Ly$\beta$ fluorescence is dominant, photons
  of this line are converted to those of neutral
  oxygen, affecting the population of the excited levels of hydrogen.
  This process would act as a cooling agent for the regions
  where the hydrogen lines are very optically thick. Therefore, it is
  a process that must be taken into account when using
  photoionization modeling to describe the state of the BLR.
  Also, the \ion{O}{1} lines can be used as a reliable
  reddening indicator for the BLR if
  the contributions of the different processes
  to the line flux are known in advance.

  To the best of our knowledge, the only simultaneous detection of
  \ion{O}{1} \lb8446 and \lb11287 in an AGN was reported
  by \citet{rudy00} for I\,Zw\,1. Previously,
  \citet{rrp89} reported the detection of
  \ion{O}{1} \lb11287 for that same object. Combining previous
  measurements of \lb8446 taken from \citet{pemac85}, \citet{rrp89} found that
  the ratio of photon fluxes between these two lines was
  equal to unity,
  providing a direct confirmation that the broad permitted \ion{O}{1}
  lines observed in an AGN arise
  through Ly$\beta$ fluorescence. Interestingly, in the spectrum
  of I\,Zw\,1 recently published by \citet{rudy00},
  with a higher resolution and spectral coverage, the photon
  flux ratio between \lb11287 and \lb8446 that we
  derive from their data is 0.76. If this ratio is corrected for
  the E(B-V) of 0.16 adopted by these authors, it does not appreciably
  change, but decreases to 0.74. Thus,
  contrary to what has been claimed, Ly$\beta$ fluorescence seems
  not to be the only mechanism contributing to the production of
  \lb8446.

  We have started a program to simultaneously
  measure \ion{O}{1} \lb8446, \lb11287 and \lb13165
  in a sample of AGN. Our goal is to establish if
  there is an excess of \lb8446 emission, as in I\,Zw\,1,
  indicating that
  Ly$\beta$ fluorescence is not the only excitation process that
  may give rise to the broad permitted \ion{O}{1} lines. For
  this purpuse, near-infrared spectroscopy at
  moderate resolution (R$\sim$320 km/s) covering the spectral range
  0.8$\mu$m -- 2.4$\mu$m was obtained. By combining these observations with
  quasi-simultaneous visible
  spectroscopy and archival HST/IUE spectra, we expect to
  obtain additional information about the radiation processes
  that govern the \ion{O}{1} emission and to test the usefulness of
  the \ion{O}{1} lines as a reddening indicator for the BLR. The
  observations are presented in \S 2, a review of the principal
  mechanisms for the production of the \ion{O}{1} lines is given
  in \S 3. A discussion of the results appears
  in \S 4. Our conclusions are in \S 5.

  \section{Observations}

  \subsection{The sample}

  The sample of galaxies used in this work is composed of six
  narrow-line Seyfert 1 galaxies (NLS1s) and  one classical Seyfert 1 galaxy.
  All seven objects show bright \ion{O}{1} $\lambda$8446, so it is
  expected that if Ly$\beta$ is the dominant excitation mechanism for
  the \ion{O}{1} emission, \ion{O}{1} $\lambda$11287 should also be present
  at a similar intensity. In addition, data in
  the UV and optical region are available for the selected
  sample to help to distinguish other processes that can also
  contribute to the formation of broad \ion{O}{1} lines. The
  reason for the predominance of NLS1s
  in the sample is that permitted lines
  in these objects are relatively narrow,
  minimizing the effects of blending with adjacent
  features. For example, the red wing of $\lambda$8444 is
  very close in wavelength to the two \ion{Ca}{2} lines located at
  $\lambda$8498 and $\lambda$8542. Table~\ref{basicdata} lists
  the objects selected for this study as well as their relevant
  characteristics. Except for I\,Zw\,1, for which the flux measurements
  were drawn from the literature, our own observations or archival
  data are used for the remaining objects, as described
  in the following sections.

  \subsection{Near-infrared spectroscopy}

  Near-infrared spectra providing continuous coverage
  between 0.8 $\mu$m and 2.4 $\mu$m were obtained at the
  NASA 3\,m Infrared Telescope Facility (IRTF) on 2000 October
  11 (UT) with the SpeX facility spectrometer
  \citep{rayner01}. The detector consists of a
  1024$\times$1024 ALADDIN 3 InSb array with a spatial scale
  of 0.$\arcsec$12/pixel. Simultaneous wavelength coverage
  is carried out by means of prism cross-dispersers. A
  0.8$\arcsec \times$15$\arcsec$ slit was used during the
  observations, giving a spectral resolution of 320 km/s.
  The seeing was near 1$\arcsec$ during the exposures. 
  Table~\ref{logopt} summarizes the log of these observations.

\begin{figure}
\plotone{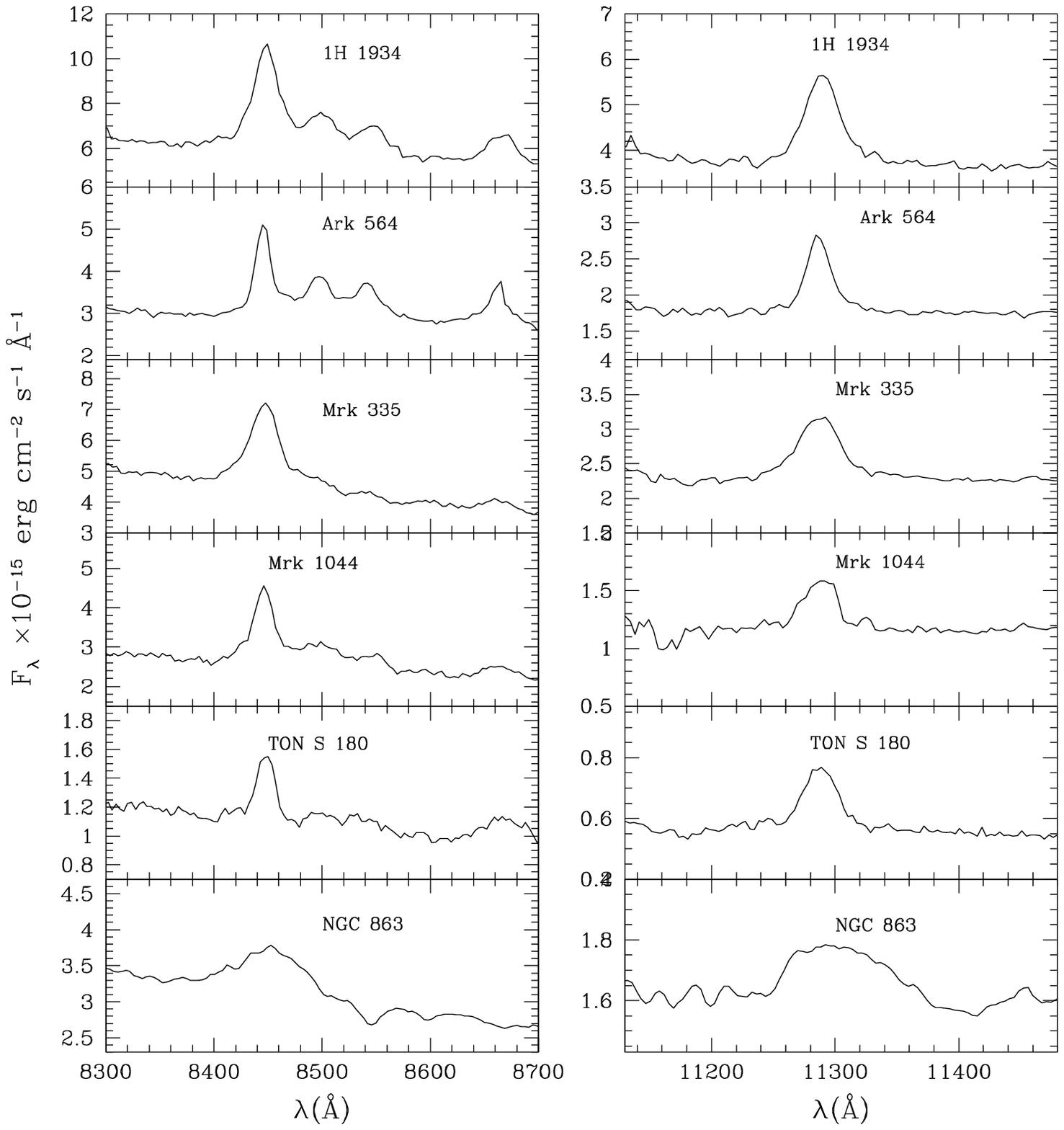}
   \caption{Observed \ion{O}{1} \lb8446 (left) and \ion{O}{1}
           \lb11287 (right) lines in rest wavelength scale.
           \label{oi_ir}}
   \end{figure}

  Details of the spectral extraction and wavelength
  calibration procedures are given in \citet{rodr01}. The
  spectral resolution was sufficiently high so that in all
  cases \ion{O}{1} $\lambda$8446 could be cleanly isolated from
  the \ion{Ca}{2} lines. \ion{O}{1} $\lambda$11287 is
  located in a region free of nearby emission features, so its
  line profile is well defined. Figure~\ref{oi_ir} shows
  the observed \ion{O}{1} lines plotted on a laboratory wavelength
  scale.


  \subsection{Optical Spectroscopy}

  Optical spectra covering simultaneously H$\alpha$ and H$\beta$
  were available for all the target objects except NGC\,863.
  Long-slit spectroscopic observations of 1H\,1934-063, Mrk\,335,
  Mrk\,1044 and TON\,S\,180 were obtained with the Complejo
  Astron\'omico El Leoncito 2.15 m telescope using a
  TEK 1024$\times$1024 CCD detector and a REOSC spectrograph.
  A 300 line mm$^{-1}$ grating with blaze angle near 5500 \AA\
  and a spatial scale of 0.$\arcsec$95 pixel$^{-1}$ was used,
  giving an instrumental resolution $\sim$8 \AA. The slit, with a width
  of 2.$\arcsec$5, was oriented in the east-west direction.
  Spectra were reduced with
  stardard IRAF procedures (i.e. bias subtraction and
  flat field division). The signal along the central 4
  pixels was summed to extract the spectra. Wavelength
  calibration was carried out using HeNeAr arc lamps, with typical
  errors of less than 0.1\AA. Standard stars from \citet{stb83}
  were used for flux calibration. In addition,
  a medium-resolution spectrum ($\sim$ 3 \AA) of
  1H\,1934-063, taken with that same telescope and covering the
  wavelength interval 6050$-$7600 \AA, was also
  available. In the remainder of this
  text, we will refer to the spectrum of 1H\,1934-063 taken
  with the 300 line mm$^{-1}$ grating as the low-resolution
  spectrum and the one taken with the 600 mm$^{-1}$ grating
  as the high-resolution spectrum.

  Ark\,564 was observed in 1996 with the {\it HST} using the Faint
  Object Spectrograph (FOS) with a circular aperture 1$\arcsec$
  in diameter.  For this object, we retrieved the calibrated
  files from the {\it HST} data archive to obtain the
  wavelength, absolute flux and error flags. A log of the
  optical observations is listed in Table~\ref{logopt}.

  \subsection{Ultraviolet Spectroscopy}

  Ultraviolet spectra observed by the {\it IUE} and the {\it HST}
  (FOS and STIS), in the wavelegth region around \ion{O}{1}  $\lambda$1304
  are available for all objects of the sample. The {\it IUE} spectra were
  re-extracted using the IUE New Spectral Image Processing System
  (NEWSIPS). When more than one observation for a given
  object was available, they were combined in order to produce
  a single, averaged spectrum. FOS and/or STIS
  nonproprietary archival calibrated files for Ark\,564,
  Mrk\,335 and TON\,S\,180 were retrieved from the {\it HST}
  data archive. Ark\,564 has been the subject of a
  monitoring campaing using STIS during the year 2000, so
  we averaged calibrated archival files taken between 
  May and July 2000. We compared the resulting
  spectrum with the FOS spectrum taken for this same
  object in 1996 and no significant difference was found
  in the region around $\lambda$1304. However, we preferred
  to work with the STIS spectrum because of the better
  resolution and S/N, which allowed us to separate, at a higher confidence
  level, the flux of \ion{O}{1}  $\lambda$1304 from that of
  \ion{Si}{2} \lb1306. Table~\ref{loguv} shows the details of
  all UV observations used in this work.

\begin{figure}
\plotone{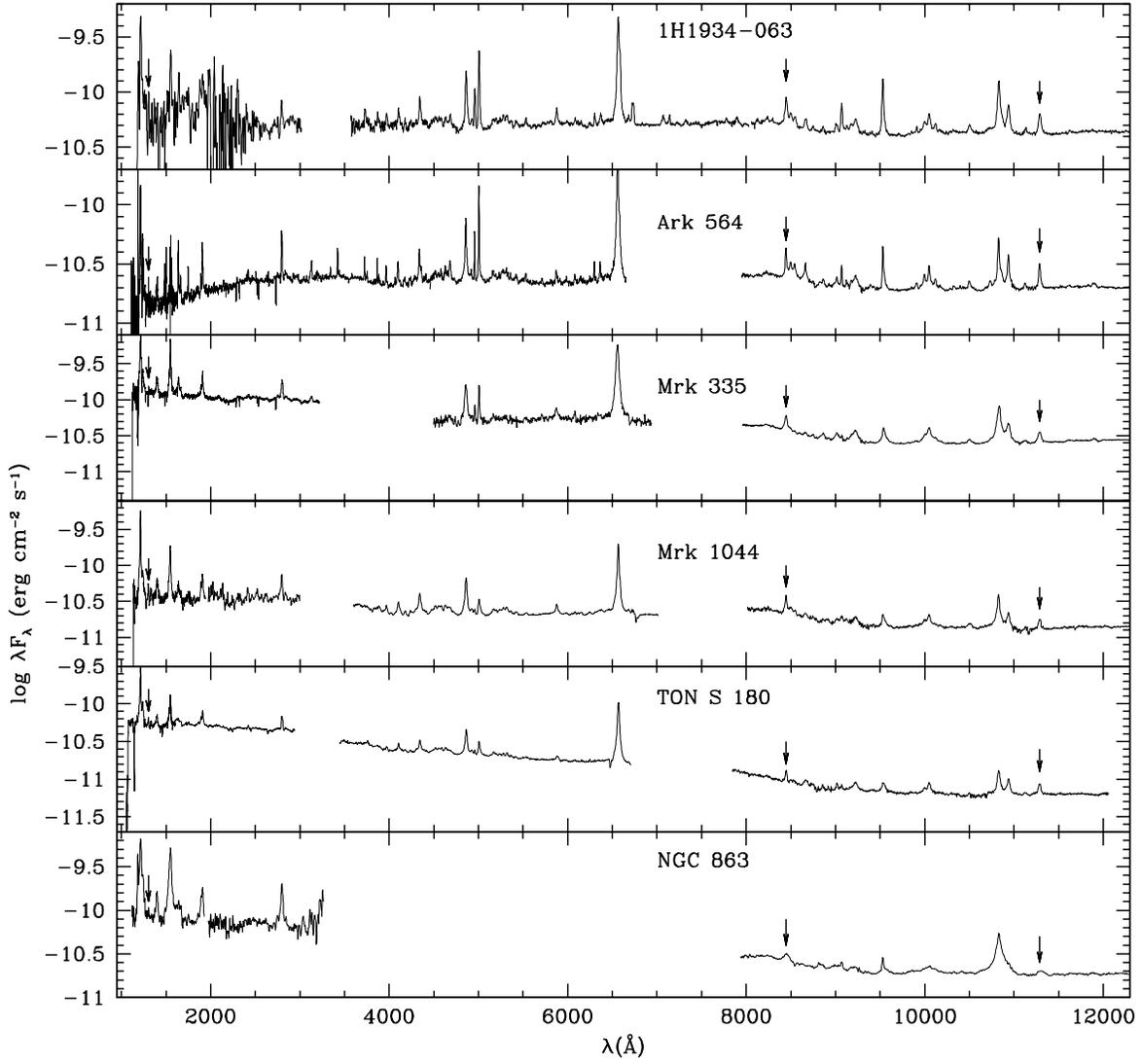}
   \caption{Composite UV, optical and near-IR spectra of
                 the galaxy sample. The arrows mark the position
                 of the \ion{O}{1} lines at $\lambda$1304, $\lambda$8446
                 and $\lambda$11287. \label{composite}}
     \end{figure}

  Figure~\ref{composite} show the composite near-IR,
  optical and UV spectra for the galaxy sample. The arrows
  mark the position of the strongest \ion{O}{1} lines that are
  of interest here.

  \subsection{Further considerations}

  Before any analysis, the spectra
  were corrected for Galactic reddening using the values
  reported in Table~\ref{basicdata}. Although different
  apertures were used during the observations, this has little
  effect on the results reported here, since our analysis
  is restricted to lines
  emitted exclusively by the BLR. Table~\ref{fluxes} lists
  the galaxy sample emission line fluxes that will be used
  throughout this paper. The erros quoted
  reflect solely the uncertainty in the placement of the 
  continuum, S/N around the line of interest and erros flags 
  of the archival data and are 2$\sigma$ significative. 
  Fluxes were measured by fitting Gaussians
  to each emission feature. In all cases, \ion{O}{1} \lb11287 and
  \ion{O}{1} \lb8446 were well described by a single Gaussian
  of similar width. In the optical region, if
  [\ion{N}{2}] \lb\lb6548 and 6583 were present, their contribution was
  always subtracted before deriving the flux for H$\alpha$. The
  strong \ion{Fe}{2} emission that contaminates the
  wings of H$\beta$ and \ion{He}{2} \lb4686 was subracted
  before measuring the flux of these two lines. For this step,
  the method used was that described by \citet{bogr92}, 
  consisting of modeling the \ion{Fe}{2}
  lines using a \ion{Fe}{2} template from I\,Zw1. 

  The greatest uncertainty comes from the measurement of
  \ion{O}{1} \lb1304, mainly in {\it IUE} data. This
  line (actually a triplet at \lb1302.17, \lb1304.86,
  and \lb1306.0) is severely blended with \ion{Si}{2}
  \lb\lb1304,1309. In order to separate the \ion{O}{1}
  contribution, we used the prescription described in
  \citet{laor97}. This consists of assuming a
  \ion{Si}{2} \lb1309/\lb1304 ratio of either 2 (optically
  thin case) or 1 (optically thick case). This method
  can be applied only if \ion{Si}{2} \lb1309 is
  clearly detected, as is the case for Ark\,564, Mrk\,335
  and TON\,S\,180. As in \citet{laor97}, we
  assumed an optically thick \ion{Si}{2} doublet given
  that for all objects of our sample the \ion{Mg}{2}
  ratio \lb2796/2804 falls between 2 and 1. In
  1H\,1934-063, Mrk\,1044 and NGC\,863, the deblending procedure
  is more uncertain because of the much lower {\it IUE}
  spectral resolution. For these objects, we assumed that
  75\% of the flux in the \lb1304 feature was
  due to \ion{O}{1}. That was the average proportion found
  in the three objects where the deblending was possible.

  \subsubsection{Variability} \label{varia}
  
  Another source of error that may affect our results
  when combining data from different spectral
  regions, taken at different dates, is variability. 
  It is widely known that Seyfert 1
  galaxies are highly variable in the UV and optical 
  continuum and broad emission
  lines. However, there is little information in
  the literature about optical and/or UV variability
  in NLS1s, even though they show the fastest variations
  in the soft and hard X-ray bands. \citet{gist96} 
  systematically monitored a sample of
  12 NLS1 during a period of one year and found that 10 of them showed
  significant variations in the optical continuum and
  permitted lines. Ark\,564 and Mrk\,1044, two of the
  objects from our sample included in
  Giannuzzo \& Stirpe's study, showed variations
  of $\sim$9\% and $\sim$10\%, respectively, in 
  permitted lines H$\alpha$ and H$\beta$. More recent data for Ark\,564
  derived from a two-year optical monitoring program \citep{shemmer01}
  showed that H$\beta$ exhibited only minor variability ($\sim$3\%)
  and that H$\alpha$ did not vary significantly. In the UV
  region, \citet{collier01}, using STIS data taken from an intensive
  two-month monitoring campaing, found flux variations of about
  1\% amplitude in the Ly$\alpha$ emission line. Moreover, they
  checked archival {\it IUE} observations of Ark 564 taken in
  Jan 1984 and compared these data with that of STIS. They 
  found that the continuum and emission line fluxes are in
  qualitative agreement in spite of the long span of the 
  observations. 
 
  For TON S 180, we have compared archival {\it IUE} spectra taken
  in 1992 and 1993 with that of STIS used in the present work.
  Although the continuum from IUE presents a
  shift towards larger flux values (in a constant amount) relative
  to the STIS spectrum, the line fluxes shows very
  little difference (by less than a factor of 10\%) between the 
  two set of observations. In addition, our flux measurement for 
  H$\beta$ taken in 2000 is in perfect agreement with that reported by 
  \citet{winkler92}, ruling out significant variations for that
  object.

  Mrk\,335 was monitored continuously in the optical region during a
  seven year campaing \citep{kassebaum97}. This NLS1 shows significant
  variability in the continuum as well as in the emission lines H$\beta$
  and \ion{He}{2} $\lambda$4686 with RMS fluctuation amplitude 
  relative to the mean spectrum of 8.5\%, 7\% and 18.8\%,
  respectively. No UV variability data for this object have been 
  reported. For 1H\,1934-063 and NGC\,863 no information about variability
  in the optical/UV region is available in the literature. 
  Optical follow-up taken for other NLS1s not included in this work, 
  such as NGC\,4051 \citep{peter00} during a three-year campaing, 
  shows RMS fluctuation amplitude relative to the mean spectrum
  of 9.5\% in H$\beta$. 

  From above, it can be seen that variability has little effect 
  on our data. Even adopting a conservative error of 10\%
  when combining UV/near-IR or UV/optical measurements,
  that estimate is within the uncertainties due to the
  deblending and S/N of the UV lines. Moreover, due to the
  fact that the \ion{O}{1} lines are formed in the outermost
  portion of the BLR (see \citet{rodr01}), it is expected
  that their variability amplitude be 
  even smaller that that found for other broad line features.

  \section{Excitation mechanisms of the \ion{O}{1} lines}

  \ion{O}{1} $\lambda$8446 can be produced by four known mechanisms:
  recombination, collisional excitation, continuum
  fluorescence, and Ly$\beta$ fluorescence. \citet{gran80},
  \citet{rrp89}, and \citet{rudy91} gave a
  complete review of these four processes. For this reason,
  here we will only focus on the resulting \ion{O}{1} emission
  lines arising from each of them.

\begin{figure}
\plotone{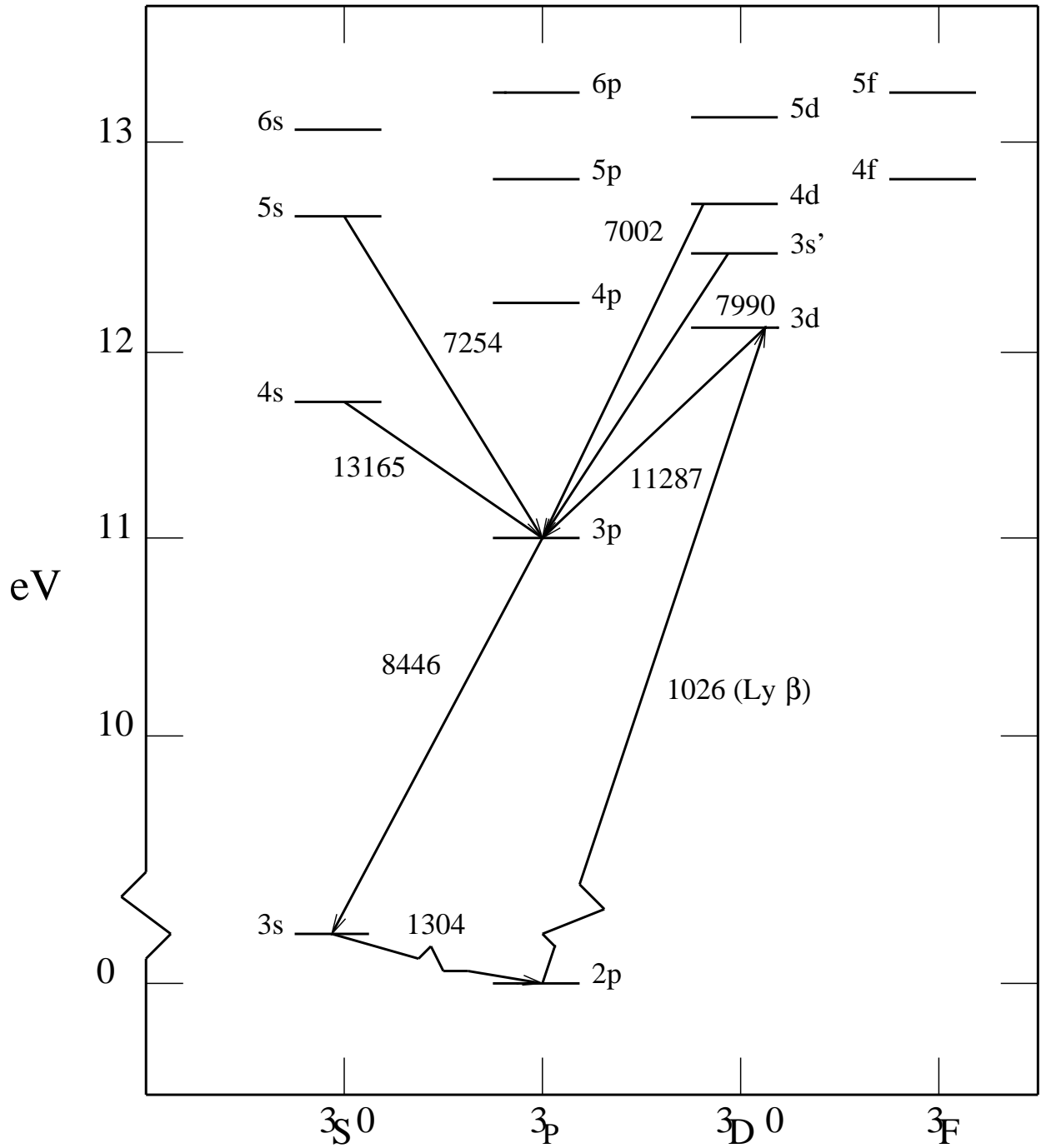}
     \caption{Partial Grotrian diagram for the triplet levels
                 of \ion{O}{1}, showing the transitions excited by Ly$\beta$.
                \label{oiniveis}}
     \end{figure}

  Besides \ion{O}{1} $\lambda$8446, recombination and collisional
  excitation also forms \ion{O}{1}
  $\lambda$7774,  while continuum fluorescence forms \ion{O}{1}
  $\lambda$7002, $\lambda$7524 and $\lambda$13165.
  Ly$\beta$ fluorescence, also known as Bowen Fluorescence,
  does not produce neither of the last four lines; their
  absence is a good indicator of the predominance of this
  mechanism. Ly$\beta$ fluorescence
  takes advantage of the
  near coincidence of the energy levels of Ly$\beta$
  (1025.72 \AA) and the 2$p~^{3}P-3d~^{3}D^{0}$
  transition of \ion{O}{1} at 1025.77, as can be seen in
  Figure~\ref{oiniveis}.
  At 10$^{4}$ K, a typical temperature within the BLR, the
  difference in wavelength falls within the Doppler core
  of Ly$\beta$. When excitation to the 3$d ^{3}D^{0}$ occurs,
  the \ion{O}{1} electron may return directly to the ground state or
  otherwise suffer a series of transitions producing photons
  at $\lambda$11287, $\lambda$8446, and $\lambda$1304
  before arriving at the ground state.


  Until now, Ly$\beta$ fluorescence has been the favoured
  mechanism for the production of $\lambda$8446 in
  AGN \citep{oksh76, gran80, morward89, rrp89}.
  The reason for this conclusion has been the apparent
  lack of detection of additional \ion{O}{1} transitions
  such as $\lambda$7002 and  $\lambda$7774. However,
  weak broad features in low resolution spectra tend to be
  smeared out when the nuclear continuum is strong and the
  S/N low. A direct
  confirmation of the predominance of the Bowen fluorescence
  can be obtained by measuring the photon flux ratio between
  $\lambda$11287 and $\lambda$8446, which must be unity (assuming no
  internal reddening). Lower values would indicate the
  presence of one of the other three mechanisms mentioned above,
  or some combination of these. This test has been elusive in
  AGNs for many years because of the lack of suitable observations
  and dectectors in the near-IR. To the best of
  our knowledge, the only detection of \ion{O}{1} $\lambda$11287
  to date in an AGN has been for the prototype NLS1 I\,Zw\,1.

  Nonetheless, we find an inconsistency in the published
  data. \citet{rrp89}
  provided a direct confirmation of the Bowen mechanism in
  I\,Zw\,1 by deriving an \ion{O}{1} $\lambda$11287/$\lambda$8446
  photon flux ratio of 1.02.  Using more recent and higher quality
  data, taken at a higher resolution
  by \citet{rudy00} for the same object,
  we derive a photon flux ratio of 0.76
  using the emission line flux reported by those
  authors in their Table 1. \citet{rrp89}
  combine data taken at different times with a
  different instrumental setup and resolution,
  while in \citet{rudy00},
  both lines are observed simultaneously, with a superior
  quality spectrograph. Our estimate
  implies that in I\,Zw1, contrary to what was previously
  found, \ion{O}{1} $\lambda$8446 is formed by more than one
  process.

  Table~\ref{ratios} lists, in column 2,
  the \ion{O}{1} $\lambda$11287/$\lambda$8446 (hereafter
  ROI$_{\rm IR}$) flux ratio
  and, in column 3, the photon flux ratio of these two lines,
  derived for the objects in our sample. Only in TON\,S\,180
  and Ark\,564, does the observational evidence points
  to Ly$\beta$ fluorescence as the dominant mechanism for
  the production of $\lambda$8446 since the photon flux
  ratio is near unity. But in Mrk\,1044, for example, Ly$\beta$
  fluorescence accounts for only $\sim$40\% of the
  observed $\lambda$8446. Evidently, additional excitation
  mechanisms that enhance \ion{O}{1} \lb8446 relative to
  \lb11287 are at work.
  Reddening cannot be invoked to explain this discrepancy
  because its net effect is the opposite: to reduce
  \ion{O}{1} \lb8446 relative to \lb11287. If the line ratios
  in Table~\ref{ratios} are affected by reddening, after
  correcting for this effect they will appear even smaller.

  It may be argued that the deficit of \lb11287 photons
  can be due to strong, sharp, atmospheric absortion
  lines not readily apparent in low-resolution spectra.
  If such a line falls near \lb11287 (in the Earth-frame
  wavelength), it can substantially affect the measured 
  flux of the latter. \citet{kingdon91} addressed this 
  possibility in his study of \ion{He}{1} \lb10830 in
  planetary nebulae and concluded that this effect could 
  reduce the intensity of \ion{He}{1} \lb10830 by as
  much as 20\%$-$25\%. Nonetheless, as he stressed,
  the effect can be dramatic in sharp-lined objects
  such as HII regions and planetary nebulae. For the
  particular case of \ion{O}{1} \lb11287,
  this line is spread over at least 21
  pixels (the narrowest full width at zero-intensity measured 
  for \ion{O}{1} \lb11287) and located in 
  different regions of the CCD due to the different redshifts 
  of the objects. 

\begin{figure}
\plotone{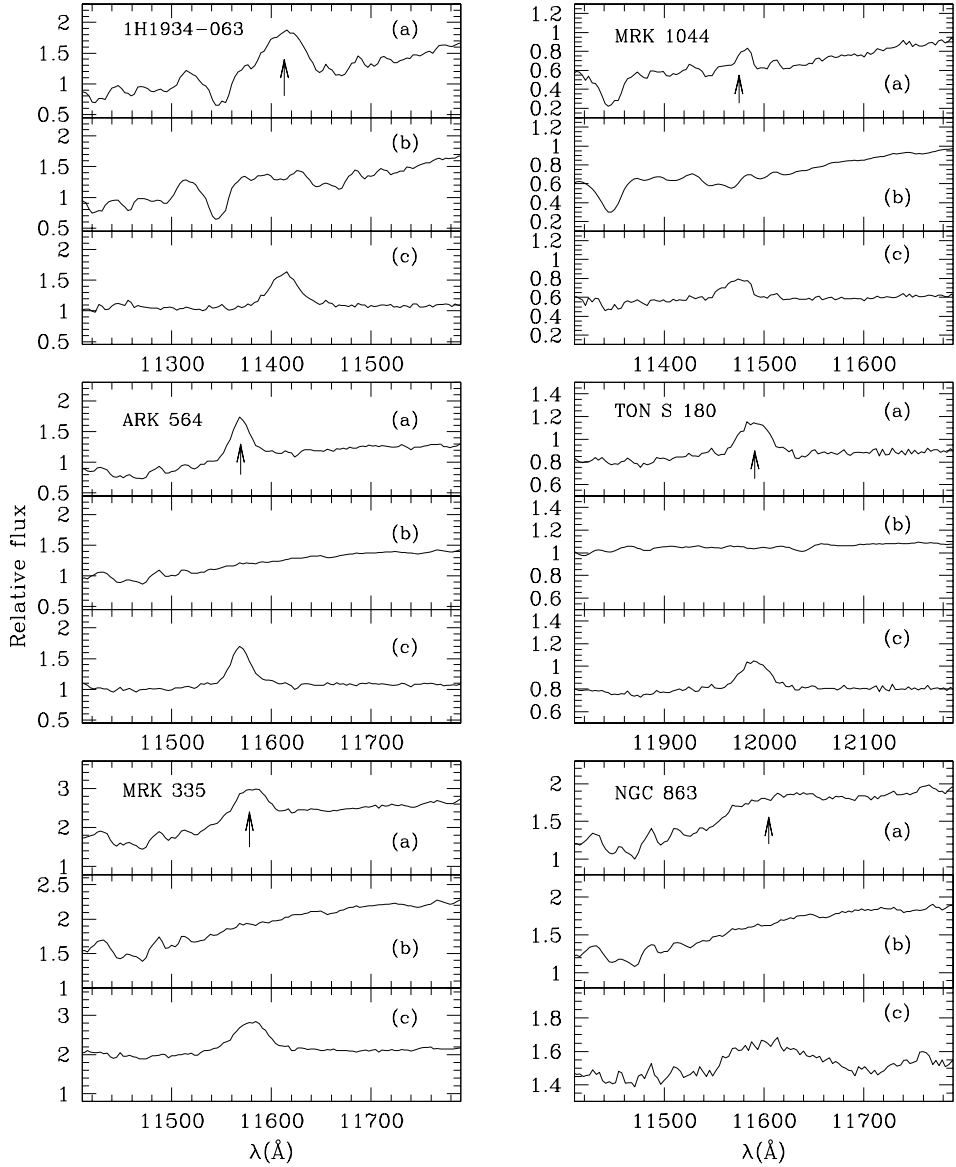}
     \caption{Effect of the atmospheric absoprtion on \ion{O}{1}
      $\lambda$11287. For each galaxy, panel (a) shows the 
      observed spectrum in the region around that line, in Earth-frame 
      wavelength coordinates. The arrow marks the centroid of 
      $\lambda$11287. In (b) 
      it is plotted a sample of the atmospheric transmission in the 
      same spectral region of (a), obtained from the spectrum
      of a telluric standard near the object's position and observed at 
      a similar airmass. Panel (c) shows the spectrum of the object 
      divided by that of the star. Note that most conspicous atmospheric 
      features dissapeared.
                \label{stardiv}}
     \end{figure}

  We consider that for \ion{O}{1} \lb11287, the most 
  serious effect is not due due to strong, sharp, atmospheric absortion
  lines but broad atmospheric bands located blueward 
  of \lb11287. If they
  are not adequately cancelled out after dividing the object's spectrum
  by that of the telluric star, the shape and flux of the line 
  can be seriously affected. Figure~\ref{stardiv} shows, for every object, 
  the observed spectrum, the spectrum of the telluric star 
  observed at similar airmass, and the resulting 
  spectrum after the division, in the region around \ion{O}{1} \lb11287.
  Note that in all cases the atmospheric features were
  cleanly removed. Although some residuals may be present, by no way 
  they do not affect the flux of \lb11287. Also, the fact that the 
  shape and width of \lb11287 is essentially the same as that of \lb8446
  (see \citet{rodr01}), confirms that the deficit of photons
  in \ion{O}{1} \lb11287 cannot be due to this effect. Regarding
  \lb8446, the position where it falls is far from any atmospheric 
  absorption feature. The nearest one is located redward of
  9000 \AA\ (in the Earth-frame wavelength) while for the galaxy 
  with the largest redshift (TONS180), \ion{O}{1} \lb8446 is located at 
  8970 \AA.


  \section{Discussion}

  The results obtained in the last section provide,
  for the first time, strong evidence that Ly$\beta$
  fluorescence cannot be the only process responsible 
  for the observed strength of \ion{O}{1} $\lambda$8446
  in most AGNs of our sample. In order to study
  which of the remaining three mechanisms already
  mentioned (continnum fluorescence, recombination and
  collisional ionization) are also producing $\lambda$8446, we use other
  permitted lines of \ion{O}{1} as diagnostics
  tools.

  According to \citet{gran80}, continuum
  fluorescence can be probed by the presence of
  \ion{O}{1} lines such us \lb7254, \lb7002 and
  \lb13165. The later line is, in fact,
  a very good test because it should be
  one of the strongest ones, with a strength similar to that
  of \lb11287, if that process is present. Unfortunately, 
  observing \lb13165 is not always easy.
  Depending on the radial velocity of the galaxy,
  it may fall very near or within a strong atmospheric
  band located between 13300 \AA\ and 14150 \AA,
  preventing its observation.

\begin{figure}
\plotone{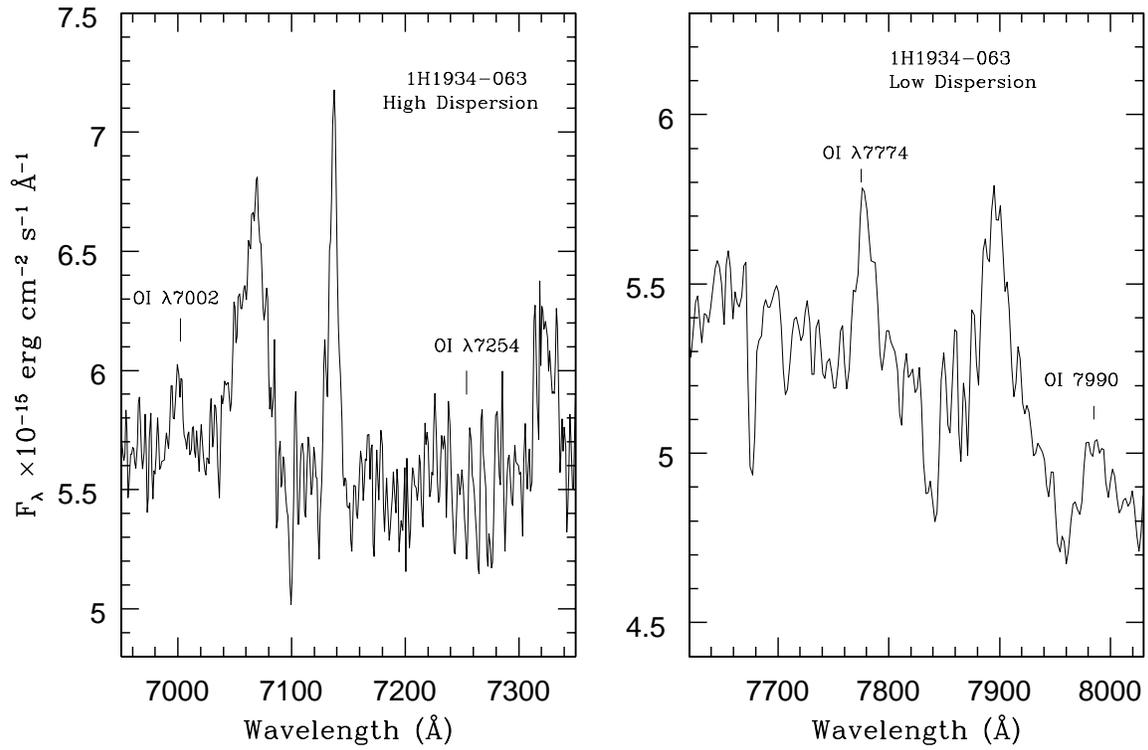}
    \caption{Other \ion{O}{1} features detected in 1H\,1934-063
            that arise from processes different to Ly$\beta$ fluorescence
            \label{oi7002}}
     \end{figure}

  We have searched carefully in the spectra for the
  presence of \ion{O}{1} \lb13165. No conclusive evidence
  of this line was found in any galaxy. Upper limits derived
  for its flux are listed in Table~\ref{fluxes}. In all cases,
  it is detected at a limit of $<$0.1F$_{\lambda 11287}$.
  In TON\,S\,180 and NGC\,863 it is not observable
  at all because it falls
  within a region of poor transmission. We also searched for
  \ion{O}{1} \lb7002 and \lb7254 in 1H\,1934-063. In the high resolution
  spectrum of this object, the former line is clearly present,
  as can be seen in Figure~\ref{oi7002}. For the latter line,
  an upper limit for its flux was derived. Table~\ref{1h1934flux} lists
  the values found.  For Ark\,564, no trace of \lb7002 can be seen in
  the spectrum published by \cite{com01}.

  The lack of additional \ion{O}{1} lines that should be observed if
  continuum fluorescence is also responsible for \lb8446
  emission and, most importantly, the very low upper limit measured
  to \ion{O}{1} \lb13165, leads us to conclude that this
  mechanism, although probably present, cannot be responsible for
  the observed excess of \ion{O}{1} \lb8446 emission.

  Recombination is another excitation mechanism that
  could help to explain the strength of \ion{O}{1} \lb8446.
  However, this process produces numerous lines located
  in the visible region, arising from triplet and quintet
  configurations, in proportions corresponding to the ratio of
  their statistical weights \citep{grandi75}. This
  implies that some of the additional \ion{O}{1} lines should
  have comparable or larger intensities than \ion{O}{1} \lb8446.
  In 1H\,1934-063, for which we have complete wavelength
  coverage from 0.37$\mu$m to 2.4$\mu$m, the only quintet
  line of \ion{O}{1} detected at 3$\sigma$ level is \lb7774
  (3$s ^{\rm 5}S - 3p ^{\rm 5}P$). This line is the quintet
  counterpart of \ion{O}{1} \lb8446. Its predicted relative
  intensity, assuming recombination as the excitation mechanism,
  should be \lb7774/\lb8446 $\sim$1.7 \citep{gran80}.
  That value is 40 times higher than the
  observed one (\lb7774/\lb8446 $\sim$0.07, see
  Table~\ref{1h1934flux}). For this object, if we subtract
  the fraction of the \ion{O}{1} \lb8446 flux produced
  by Ly$\beta$ fluorescence, the resulting \lb7774/\lb8446
  ratio is 0.2, nearly 20 times weaker than its theoretical
  prediction.

  The other object for which observations around
  \ion{O}{1} \lb7774 are available is TON\,S\,180. The upper limit
  derived for this feature, also listed in Table~\ref{1h1934flux},
  is two orders of magnitude weaker than the theoretical prediction.
  However, it is not expected that this line be strong
  in TON\,S\,180 since Ly$\beta$ fluorescence accounts for
  the total \ion{O}{1} \lb8446 emission.

  In addition to \ion{O}{1} \lb7774, we detect in
  1H\,1934-063, \ion{O}{1} \lb7990 (see Figure~\ref{oi7002})
  and \lb6048. A blip at this later position is also observed
  in the spectrum of Ark\,564. Nonetheless, their intensities
  are always a few percent of \ion{O}{1} \lb8446. For this
  reason, we conclude that recombination is certainly present
  but it cannot explain alone the excess of \lb8446 observed in
  most of the objects.

  The only possibility that remains to be explored is
  collisional excitation by electrons. \citet{gran80}
  states that if this mechanism is operating, the
  predicted \ion{O}{1} \lb7774/\lb8446 ratio would be 0.3.
  That value is near the observed ratio (0.2)
  derived in 1H\,1934-063 after removing the contribution 
  of Ly$\beta$ fluorescence for \ion{O}{1} \lb8446. 
  The lack of observations around \lb7774 for
  the remaining objects prevent us from drawing definite
  conclusions. However, our data clearly show that
  L$\beta$ fluorescence cannot be considered any longer as
  the single dominant mechanism for the production of the observed
  \ion{O}{1} lines in AGN. For 1H\,1934-063, for instance, 
  collisions may contribute up to 35\% of the observed
  \ion{O}{1} \lb8446 flux. 

  \subsection{An optically thick \lb1304 line?}

  If collisional excitation is responsible for the excess of
  \ion{O}{1} \lb8446 emission, in order to this process
  be efficient the population of the level 3$s^{3}S^{0}$ must
  be significant, i.e. above the Boltzman population. This
  requires high optical depths in the \lb1304 line. \citet{kwkro81} 
  state that with increasing column density,
  trapping of \lb1304 becomes important in sustaining the
  population in 3$s^{3}S^{0}$, and \lb8446 becomes optically
  thick. Trapping of \lb8446, in turn, builds up the population
  in 3$p^{3}P$, eventually leading to trapping of \lb11287.
  As a result, the growth of these lines is subsequently quenched;
  once \lb11287 is optically thick, all three lines grow only
  logarithmically with optical depth. This possibility
  has been examined by \citet{feper89},
  who found that the very high column densities ($>$10$^{24.5}$
  cm$^{-2}$) necessary to produce \ion{Ca}{2} can produce
  large optical depths in the \ion{O}{1} lines. In fact,
  very recently, \citet{rodr01} found
  that the \ion{O}{1}, \ion{Ca}{2} and \ion{Fe}{2} lines
  have very similiar emission line profiles, both in form
  and width. This is interpreted as a co-spatial
  \ion{Ca}{2} and \ion{O}{1} emitting regions, which probably share similar
  physical conditions.

  An indication that \lb8446 is thick would be branching from
  its upper level to 3$^{5}S$, by the \ion{O}{1}] \lb6726 line
  \citep{rudy91}. This line is very close to
  [\ion{S}{2}] \lb6717,6731, a commom doublet in Seyfert 1
  galaxies, so if this pair of lines is present, it
  may mask the observation of \ion{O}{1}] \lb6726. We have
  searched our spectra in order to find evidences for
  \lb6726. Table~\ref{1h1934flux} list the derived upper
  limits. In none of the four objects where this measurement was
  possible was the \lb6726 line present at
  a level greater than
  0.04$I$(\lb8446). We conclude that the intensity of \lb8446
  is not significantly altered by optical depth effects.
  Probably, after the upper 3$p^{3}P$ level is
  significantly populated, the electrons are rapidly
  promoted to the upper $^{3}S^{0}$ and $^{3}D^{0}$
  levels because of their close proximity in energy
  (less than 2 eV), producing the additional \ion{O}{1}
  lines detected and increasing the flux of \lb8446.

  \subsection{\ion{O}{1} as a reliable reddening indicator for
  the BLR}

  According to the Grotrian diagram of Figure~\ref{oiniveis},
  independent of the excitation mechanism, every
  $\lambda$8446 photon should produce a $\lambda$1304 photon.
  Theoretically, the photon flux ratio
  $\lambda$1304/$\lambda$8446 (ROI$_{\rm uv}$, hereafter) must
  be equal to unity except when reddening is present, in which case,
  it should be smaller. \citet{kwkro81} found that the intrinsic ROI$_{\rm uv}$
  can vary from 1 to 0.63 even in the absence of reddening due to 
  Balmer continuum absorption of \lb1304
  photons and the production of \lb8446 by collisional excitation.
  \citet{grandi83} describes three additional
  proccesses that may destruct \lb1304: {\it (i)} $\lambda$1304 photons can
  photoionize \ion{H}{1} atoms that exists in the $n=2$ state;
  {\it (ii)} collisional de-excitation can destroy $\lambda$1304
  photons; {\it (iii)} the upper term of $\lambda$1304 (3$s~^{\rm 3}S^{o}$)
  can decay to the metastable terms of the ground configuration
  via the semi-forbidden lines \ion{O}{1}] \lb1641 and \lb2324
  ($2p^{4}~^{\rm 1}D$ - $3s~^{3}S^{o}$ and $2p^{4}~^{1}S$ - $3s~^{3}S^{o}$,
  respectively). Grandi's calculations show that up to half of
  the $\lambda$1304 photons can be converted to \ion{O}{1}] \lb1641 before
  leaving the emission cloud.

  Our data offers, for the first time, the
  possibility of evaluating, at least to first order, how efficient
  the destruction of the \lb1304 photons can be. Column 5 of
  Table~\ref{ratios} lists the ROI$_{\rm uv}$ photon ratio
  derived for the galaxies. It was determined assuming
  an intrinsic flux ratio of 6.5. Note that the uncertainties
  in this quantity is, in most cases, larger that the estimated
  error due to variability (see \S ~\ref{varia}), meaning that
  the results are not at all affected by this effect. 
  Except TON\,S\,180, all objects show flux ratios significantly
  below the lowest theoretical limit derived by \citet{kwkro81}. 
  Even if in all these objects the \lb1304 photons
  are severely affected by the destruction mechanisms described above,
  they are not sufficient to explain the break of the one-to-one
  photon relation between \lb8446 and \lb1304.  TON\,S\,180 is
  particularly interesting in this respect. We had already seen
  that it was the only galaxy where the ROI$_{\rm ir}$ ratio
  pointed to Ly$\beta$ fluorescence as the dominant mechanism
  for producing \lb8446. The ROI$_{\rm uv}$ ratio not only
  confirms this result but also indicates that the physical conditions
  of the \ion{O}{1} gas is different for this  object. 
  Apparently, the optical depth effects that may affect the 
  \ion{O}{1} lines do not apply here.

  The departure from the one-to-one photon relation can be interpreted
  as due only to reddening or the combined effects of reddening and
  the destruction mechanisms of \lb1304. Columns 2 and 3 of Table~\ref{redd}
  list the E(B-V) inferred from the observed flux ratio. Values 
  listed in columns 2
  and 3 assume an intrinsic ROI$_{\rm uv}$ photon ratio of 1 and 1.6, 
  respectively (which translates into a flux ratio of 6.5 and 4). 
  The Galactic extinction 
  curve of \citet{ccm89} were employed in this calculation.

  The E(B-V) derived from the \lb1304/\lb8446
  ratio points toward low values of reddening in the BLR of
  Seyfert 1 galaxies, except in Ark\,564, which seems to
  be affected by a moderate amount of extinction. TON\,S\,180,
  on the other hand, shows evidence of being reddenning free.

  In order to compare the E(B-V) derived from the \ion{O}{1}
  lines with that predicted from other indicators, we have
  also calculated the reddening using the Balmer decrement
  H$\alpha$/H$\beta$ and the \ion{He}{2} \lb1640/\lb4686 ratio.
  The results are shown in columns 4 and 5 of Table~\ref{redd},
  respectively. A ``N.A.'' entry means that that particular
  ratio was not available for measurement in the spectrum and the
  lack of any value means that the calculated ratio gave
  unphysically results (the theoretical values adopted were
   3.1 for the Balmer lines and 7.2 for the \ion{He}{2} lines). Although
  the results of Table~\ref{redd} are approximate, mainly because
  of the uncertainy involved in the knowledge of the intrinsic
  ratio of the Balmer decrement and the \ion{He}{2} lines
  for the BLR, it is encouraging to see that the agreement
  between the different reddening indicators for the same object,
  are good to within 0.1 mag of uncertainty. In particular, for
  TON\,S\,180, and Mrk\,1044, the three indicators agree to very
  little or no reddening at all, while for Ark\,564, they clearly
  confirm the presence of dust. The highest discrepancy is
  observed in 1H\,1934-063. This result can be explained if we
  consider that the S/N of the UV spectrum is very low and just
  upper limits to the flux of the lines were derived.

  It is also interesting to see that the values of E(B-V)
  derived assuming an intrinsic \lb1304/\lb8446 flux ratio
  of 4 are closer to the ones derives using other indicators
  than if we assume an intrinsic ratio of 6.5 (except for
  TON\,S\,180). This provides strong observational support to
  the hypothesis of \lb1304 photons being destroyed by
  Balmer continuum absorption and the formation of
  \lb8446 by collisional excitation. Although detailed modelling
  of the BLR is beyond the scope of this paper, it would
  be important to estimate how much the intrinsic
  \lb11287/\lb8446 ratio is affected by this latter effect.
  The data provided here is a good starting point to
  investigate the physics of the outer BLR.

  \section{Summary and Conclusions}

  Near-infrared, optical and UV spectroscopy has been used
  to study the excitation mechanisms leading to the
  formation of permitted \ion{O}{1} lines in Seyfert 1 galaxies.
  From the observed \lb11287/\lb8446 photon flux ratio,
  we found that in only one out of seven
  galaxies, L$\beta$ fluorescence is the single dominant process
  in the formation of \ion{O}{1} \lb8446. This
  result clearly contradicts previous assumptions that
  L$\beta$ fluorescence is the only contributor for the
  formation of the \ion{O}{1} lines in those objects. Continuum fluorescence
  is discarded as an additional mechanism due to the very
  low \ion{O}{1} \lb8446/\lb13165 flux ratio measurement as well as
  the absence of other
  emission lines that should be present in the optical region.
  Recombination is not negligible, although its
  effects on the flux of \lb8446 seems to be no larger than
  a few percent. Collisional excitation offers a
  plausible explanation to enhance the strength of the \ion{O}{1}
  \lb8446 relative to that of \lb11287, leading to
  the strong deviation from unity observed
  in the \lb11287/\lb8446 ratio. This conclusion is drawn
  from the presence of \ion{O}{1} \lb7774 in 1H\,1934-063.
  The strength of this line, relative to that of \lb8446,
  agrees with theoretical predictions assuming collisional
  excitation as responsible for the formation of \lb7774.
  In that object, up to 35\% of the observed \lb8446 flux may
  arise through collisional pumping. 
  Collisional excitation is also favoured if \lb1304 is
  optically thick, as is suggested by the presence of strong
  \ion{Ca}{2} lines, which requires high column densities
  for their formation.

  In the absence of reddening, the \lb1304/\lb8446 photon
  flux ratio may depart from its intrinsic value (from 1 to
  1.6) because of several mechanisms that destroy
  the \lb1304 photons, as well as optical depth effects
  that favor collisional excitation of \lb8446. The agreement
  between the E(B-V) of the BLR derived using different
  reddening indicators (among them, the ratio \lb1304/\lb8446)
  is improved if an intrinsic flux ratio \lb1304/\lb8446 of 4
  is assumed, meaning that destruction effects for the \lb1304
  line are, in fact, present.  Our results points towards low to 
  moderate reddening for the BLR.

  \acknowledgements

  We thank the IRTF staff, support scientist Bobby Bus,
  and telescope operator Dave Griep for contributing to a
  productive observing run. Mike Cushing provided patient
  assistance with the xspextool software. The authors thank
  the sugestions of an anonymous referee, which help to improve
  this paper. This research has
  been supported by the Funda\c c\~ao de Amparo a Pesquisa
  do Estado de S\~ao Paulo (FAPESP) to ARA, PRONEX grant
  662175/1996-4 to SMV and ARA and PRONEX grant 7697100300 to
  MGP and ARA.

  \clearpage

   \clearpage

   \clearpage

   \clearpage

   \clearpage

   \clearpage

  \begin{deluxetable}{lcccc}
  \tablecaption{Sample characteristics. \label{basicdata}}
  \tablewidth{0pt}  
  \tablehead{
  \colhead{Galaxy} & \colhead{z} & \colhead{M$_{\rm v}$\tablenotemark{a}} & 
  \colhead{A$_{\rm v}$\tablenotemark{b}} & \colhead{Type}}
  \startdata
  1H\,1934-063       & 0.01059 &  -19.04   & 0.972 & NLS1  \\
  Ark\,564          & 0.02468 &  -20.42   & 0.198 & NLS1  \\
  Mrk\,335          & 0.02578 &  -21.32   & 0.118 & NLS1  \\
  I\,Zw\,1          & 0.06114 &  -22.58   & 0.214 & NLS1 \\
  Mrk\,1044         & 0.01645 &  -18.84   & 0.113 & NLS1 \\
  TON\,S\,180       & 0.06198 &  -22.57   & 0.047 & NLS1 \\
  NGC\,863          & 0.02638 &  -21.27   & 0.124 & Seyfert 1 \\
  \enddata
  \tablenotetext{a}{A value of H$_{\rm o}$ = 75 km s$^{-1}$ Mpc$^{-1}$ was assumed.}
  \tablenotetext{b}{Galactic extinction \citep{sfd98}}
  \end{deluxetable}

  \clearpage

\begin{deluxetable}{llcccc}
\tablecaption{Log. of near-IR and optical observations. \label{logopt}}
\tablewidth{0pt}
\tablehead{
\colhead{} &  \colhead{Instr./Obs} &  \colhead{grating} &
\colhead{Coverage} &  \colhead{Exp. Time} &   \colhead{Date of Obs.}\\
\colhead{Galaxy}  &   & \colhead{(l/mm)}  &  \colhead{(\AA)} & \colhead{(sec)} &
}
\startdata
1\,H1934-063 &  IRTF         & \nodata & 8000 $-$ 24000 &  2700  & 2000 Oct 11 \\
           &  CASLEO/REOSC & 600   &  6050 $-$ 7600 &    3600  & 1996 Aug 10 \\
           &  CASLEO/REOSC & 300   &  3700 $-$ 6800 &    3600  & 1997 Aug 29 \\
           &  CASLEO/REOSC & 300   &  6500 $-$ 9600 &    3600  & 1997 Aug 30 \\
           &               &       &                &          &  \\
Ark\,564    &  IRTF         & \nodata & 8000 $-$ 24000 &  2400  & 2000 Oct 11 \\
           &  FOS(Y3790406T)& \nodata& 4500 $-$ 6650 &   4600  & 1996 May 23 \\
           &               &       &                &          &  \\
Mrk\,335    &  IRTF         & \nodata & 8000 $-$ 24000 &  1800  & 2000 Oct 11 \\
           &  CASLEO/REOSC & 300   &  3700 $-$ 6800 &    1200  & 2000 Dec 28 \\
           &               &       &                &          &  \\
Mrk\,1044   &  IRTF         & \nodata & 8000 $-$ 24000 &  1800  & 2000 Oct 11 \\
           &  CASLEO/REOSC & 300   &  3700 $-$ 6800 &    1200  & 2000 Dec 28 \\
           &               &       &                &          &  \\
TON\,S\,180  &  IRTF         & \nodata & 8000 $-$ 24000 &  2400  & 2000 Oct 11 \\
           &  CASLEO/REOSC & 300   &  3700 $-$ 6800 & 1200  &    2000 Dec 29 \\
           &               &       &                &          &  \\
NGC 863    &  IRTF         & \nodata & 8000 $-$ 24000 &  1800  & 2000 Oct 11 \\
\enddata
\end{deluxetable} 

\clearpage

  \begin{deluxetable}{lcccccc}
  \tablecaption{Log. of UV observations.\label{loguv}}
  \tablewidth{0pt}
  \tablehead{
    & \colhead{Inst.} & \colhead{Data Set} & \colhead{Exposure} & 
\colhead{Coverage} & \colhead{Date} \\  
\colhead{Galaxy} &   &  & \colhead{(sec)} & \colhead{(\AA)} & \colhead{(UT)}} 
\startdata
1\,H1934-063 &   IUE/SWP &   SWP31894  &  8400.00  & 1200 $-$ 2000 &  1987 Sep 21 \\
           &           &             &           &  & \\
Ark\,564    &   STIS    &   O5IT01010 &  1201.00  &  1150 $-$ 1714 & 2000 May 10 \\
           &   STIS    &   O5IT02010 &  1021.00  &  1150 $-$ 1714 & 2000 May 14 \\
           &   STIS    &   O5IT03010 &  1021.00  &  1150 $-$ 1714 & 2000 May 19 \\
           &   STIS    &   O5IT04010 &  1021.00  &  1150 $-$ 1714 & 2000 May 24 \\
           &   STIS    &   O5IT44010 &  1201.00  &  1150 $-$ 1714 & 2000 Jul 6  \\
           &   STIS    &   O5IT45010 &  1201.00  &  1150 $-$ 1714 & 2000 Jul 8  \\
           &   STIS    &   O5IT46010 &  1021.00  &  1150 $-$ 1714 & 2000 Jul 10 \\
           &           &             &           &  & \\
Mrk\,335    &   FOS     &   Y29E0202T &  1390.00  &  1150 $-$ 1600 & 1994 Dec 16 \\
           &           &             &           &  & \\
Mrk\,1044   &   IUE/SWP &   SWP56260  &  34080.00 &  1200 $-$ 2000 & 1995 Dec 7  \\
           &   IUE/SWP &   SWP56319  &  13499.00 &  1200 $-$ 2000 & 1995 Dec 20 \\
           &           &             &           &  & \\
TON\,S\,180  &   STIS     &  O58P01010  & 1260.00  &  1150 $-$ 1714 & 2000 Jan 22 \\
           &           &             &           &  & \\
NGC 863    &   IUE/SWP  &  SWP40591   & 9000.00  &  1200 $-$ 2000 & 1991 Jan 14 \\
 \enddata
\end{deluxetable}

\clearpage

  \begin{deluxetable}{lccccccc}
\tabletypesize{\small}
  \tablecaption{Fluxes of relevant emission lines\tablenotemark{a}.\label{fluxes}}
  \tablewidth{0pt}
  \tablehead{
  \colhead{Line} & \colhead{1H\,1934-063} & \colhead{Ark\,564} & 
  \colhead{Mrk\,335} & \colhead{Mrk\,1044} & \colhead{TON\,S\,180} & 
  \colhead{NGC\,863} & \colhead{I\,Zw\,1\tablenotemark{*}}}
  \startdata
  \ion{O}{1} $\lambda$1304  & $<$18.0\tablenotemark{b} & 2.00$\pm$0.20 & 10.9$\pm$1.0 & 13.6$\pm$2.00 & 
  5.74$\pm$0.7
  & 6.70$\pm$1.0 & $<$10.0\\
  \ion{He}{2} $\lambda$1640 & $<$56.5   & 11.8$\pm$0.70 & 63.0$\pm$5.0 & 24.2$\pm$3.6 &
  7.15$\pm$1.50 & 25.7$\pm$4.0 & \nodata \\
  \ion{He}{2} $\lambda$4686 & 7.00$\pm$0.80 & 3.80$\pm$0.20 & 31.0$\pm$3.0 & 4.48$\pm$0.5 &
  1.21$\pm$0.15 & \nodata & \nodata \\
  H$\beta$     & 41.83$\pm$3.0 & 22.8$\pm$0.8 & 76.5$\pm$7.0 & 26.8$\pm$2.0 & 14.5$\pm$1.0 &
  \nodata & \nodata \\
  H$\alpha$    & 129.7$\pm$5.0 & 92.3$\pm$1.0 & 320$\pm$10 & 67.4$\pm$4.0   &
  43.0$\pm$2.0 & \nodata & \nodata \\
  \ion{O}{1} $\lambda8446$  & 14.5$\pm$1.0 & 4.40$\pm$0.10 & 8.14$\pm$0.5 & 4.96$\pm$0.4 & 0.98$\pm$0.10
  & 4.01$\pm$0.30 & 8.83$\pm$1.0 \\
  \ion{O}{1} $\lambda11287$ & 6.91$\pm$0.24 & 2.70$\pm$0.07 & 3.87$\pm$0.17 & 1.52$\pm$0.17 &
  0.80$\pm$0.09 & 1.65$\pm$0.2 & 5.00$\pm$0.4  \\
  \ion{O}{1} $\lambda13165$\tablenotemark{c} & $<$0.17 & $<$0.18 &
  $<$0.20 & $<$0.22 & \nodata & \nodata & \nodata \\
  \enddata
  \tablenotetext{a}{In units of 10$^{-14}$ erg cm$^{-2}$ s$^{-1}$.}
  \tablenotetext{b}{The $<$ sign implies an upper limit.}
  \tablenotetext{c}{In TON\,S\,180 and NGC\,863 \ion{O}{1} \lb13165
  falls within an atmospheric absorption band, so it is not possible to derive
  an upper limit.}
  \tablenotetext{*}{Fluxes taken from \citet{rudy00}}
  \end{deluxetable}

  \clearpage

  \begin{deluxetable}{lccccc}
  \tablecaption{Measured ROI$_{\rm IR}$ and ROI$_{\rm uv}$
  flux and photon ratio for the galaxy sample.\label{ratios}}
  \tablewidth{0pt}
  \tablehead{
  \colhead{}  & \multicolumn{2}{c}{ROI$_{\rm IR}$} & &
  \multicolumn{2}{c}{ROI$_{\rm uv}$} \\
  \cline{2-3} \cline{5-6}
  \colhead{Galaxy}  & \colhead{Flux} & \colhead{Photon} & & 
  \colhead{Flux} & \colhead{Photon}}
  \startdata
  1H\,1934-063 &  0.48$\pm$0.04  &  0.64$\pm$0.05 &  & 1.25\tablenotemark{a} &  0.19\tablenotemark{a}  \\
  Ark\,564     &  0.61$\pm$0.02  &  0.82$\pm$0.03 &  & 0.45$\pm$0.05 &  0.07$\pm$0.01 \\
  Mrk\,335     &  0.48$\pm$0.04  &  0.64$\pm$0.05 &  & 1.33$\pm$0.15 &  0.20$\pm$0.02  \\
  Mrk\,1044    &  0.31$\pm$0.04  &  0.42$\pm$0.05 &  & 2.78$\pm$0.47 &  0.43$\pm$0.07  \\
  TON\,S\,180  &  0.81$\pm$0.12  &  1.08$\pm$0.16 &  & 5.88$\pm$0.98 &  0.91$\pm$0.15 \\
  NGC\,863     &  0.41$\pm$0.06  &  0.55$\pm$0.08 &  & 1.70$\pm$0.28 &  0.26$\pm$0.04  \\
  I\,Zw\,1     &  0.56$\pm$0.08  &  0.76$\pm$0.11 & &  1.14\tablenotemark{a} &  0.18\tablenotemark{a}  \\
  \enddata
  \tablenotetext{a}{Upper limit}
  \end{deluxetable}

  \clearpage

  \begin{deluxetable}{lccccc}
  \tablecaption{Fluxes measured to other \ion{O}{1} features\tablenotemark{a}.\label{1h1934flux}}
  \tablewidth{0pt}
  \tablehead{
  \colhead{Line} & \colhead{1H\,1934-063} & \colhead{Ark\,564} & 
  \colhead{Mrk\,335} & \colhead{Mrk\,1044} & \colhead{TON\,S\,180}}
  \startdata
  $\lambda$6046 & $<$4.6       &  $<$1.5    & $<$2.3     &  $<$2.4     & $<$0.9 \\
  $\lambda$6726 & $<$6.0       & ...     & $<$2.1     &  $<$1.3     & ... \\
  $\lambda$7002 & 4.0$\pm$0.8 & ...     & ...     &  ...     & ... \\
  $\lambda$7254 & $<$3.9       & ...     & ...     &  ...     & ... \\
  $\lambda$7774 & 10.0$\pm$0.8 & ...    & ...     &  ...     & $<$0.8 \\
  $\lambda$7990 & 5.1$\pm$0.9 & ...     & ...     &  ...     & $<$0.9 \\
  $\lambda$8446 & 145.0$\pm$10 & 44.0$\pm$1.0 & 81.4$\pm$5.0 & 49.6$\pm$5.0 & 9.8$\pm$1.0 \\
  \enddata
  \tablenotetext{a}{In units of 10$^{-15}$ erg cm$^{-2}$ s$^{-1}$. Lines
  with the $<$ sign indicate an upper limit.}
  \end{deluxetable}

  \clearpage

  \begin{deluxetable}{lccccc}
  \tablecaption{Derived E(B-V) from the \ion{O}{1} lines and other reddening
  indicators\tablenotemark{a}.
  \label{redd}}
  \tablewidth{0pt}
  \tablehead{
  \colhead{Galaxy}  & \multicolumn{2}{c}{R\,OI$_{\rm uv}$} &
  \colhead{H$\alpha$/H$\beta$} & 
  \colhead{\ion{He}{2}$\lambda$1640/$\lambda$4686} \\
  \cline{2-3}
  \colhead{} & \colhead{(6.4)} & \colhead{(4.0)} & \colhead{(3.1)} & \colhead{(7.2)}}
  \startdata
  1H\,1934-063 & 0.23 & 0.15 &  0.0 &  ... \\
  Ark\,564     & 0.36 & 0.29 &  0.26 &  0.22 \\
  Mrk\,335     & 0.21 & 0.15 &  0.30 &  0.34 \\
  Mrk\,1044    & 0.11 & 0.05 &  0.0  &  0.06 \\
  TON\,S\,180   & 0.01 & 0.00 &  0.0  &  0.05 \\
  NGC\,863     & 0.18 & 0.11 &  N.A. & N.A. \\
  I\,Zw\,1    & 0.24 & 0.17 &  N.A. & N.A. \\
  \enddata
  \tablenotetext{a}{The value in parenthesis below the reddening indicator
  corresponds to the intrinsic ratio assumed}
  \end{deluxetable}

  \end{document}